# Securing IoT Applications using Blockchain: A Survey

Sreelakshmi K. K., Ashutosh Bhatia, Ankit Agrawal
*Dept. of Computer Science*
*Birla Institute of Technology and Science, Pilani*
Rajasthan, India
{h20180130, ashutosh.bhatia, p20190021}@pilani.bits-pilani.ac.in

***Abstract*— The Internet of Things (IoT) has become a guiding technology behind automation and smart computing. One of the major concerns with the IoT systems is the lack of privacy and security preserving schemes for controlling access and ensuring the security of the data. A majority of security issues arise because of the centralized architecture of IoT systems. Another concern is the lack of proper authentication and access control schemes to moderate access to information generated by the IoT devices. So the question that arises is how to ensure the identity of the equipment or the communicating node. The answer to secure operations in a trustless environment brings us to the decentralized solution of Blockchain. A lot of research has been going on in the area of convergence of IoT and Blockchain, and it has resulted in some remarkable progress in addressing some of the significant issues in the IoT arena. This work reviews the challenges and threats in the IoT environment and how integration with Blockchain can resolve some of them.**

**Keywords—IoT, Blockchain, Security**

## I. Introduction

The rapid advancements in networking technologies have led to an increased number of devices or things being able to connect to the Internet, which forms the Internet of Things, commonly known as IoT. Some of the leading IoT applications are Smart-grid, smart-homes, Industrial IoT, smart healthcare, etc. At a high-level, a typical IoT ecosystem consists of devices like sensors that collect data, actuators, and other devices that perform control and monitoring specific to the application area, communication infrastructure guided by the network protocols and local or centralized storage (cloud) that collects data from different devices and processes it for further analysis. The data generated by the various IoT devices is characterized by its vast volume, heterogeneity, and dynamic nature. Each device in an IoT environment has a unique identifier associated with it. A review work Colakovic and Hadialic (2018) gives a detailed and technical description of IoT and its enabling technologies. To build a sustainable IoT ecosystem that can adapt and perform well in a particular application area is a challenging task. With various smart solutions using a large number of devices, the problem of maintaining security for private or user data in the centralized cloud storage is a very tedious task that needs significant attention. The stakes are high if the private data falls into the hands of malicious entities. Another issue that draws concern is the resource and energy constraints of devices, which make it challenging to run massive cryptographic algorithms that strengthen the security of the data generated. The other challenges that need to be dealt with in specific situations such as a disaster are the fault tolerance and recovery of devices located at remote locations. Also, the diverseness in the IoT application areas and the uncertainty about the technology and the solutions offered creates a lack of trust in these solutions. Thus the need for a decentralized solution to ensure the security in an IoT ecosystem is an essential requirement of any IoT application.

The concept of Blockchain has been very active in cryptographic and security since its revelation in 2008. The remarkable advancements made in the Blockchain research have made cryptocurrencies a reality using BitCoin. The BitCoin with a current value of 5431.43 USD and holding around 5.7 million blocks is continuously increasing at a rapid rate with the help of more hashing power and mining pools. Software giants like Linux Foundation are already hacking the potential of Blockchain by researching on applications like HyperLedger (Blummer et al., 2018). With the Software industry predicting a promising future for the blockchain technology, it is essential that we investigate in detail the applicability of Blockchain in IoT to address the main security issues and how far it can be successful in solving them.

When Blockchain meets IoT, some of the expected benefits with this convergence are the building of trust between entities, completeness, consistency, and integrity of stored data, immutability or tamper-proof preservation of private data, reduction in expenses and thereby cheaper solutions, enhanced security and faster processing of Big Data.

The major objectives of the review work presented in this paper are:

- To investigate the security issues and challenges currently present in IoT applications.
- To highlight the decision criteria for applying Blockchain in IoT.
- To survey the scope of solutions that can be achieved through the convergence of Blockchain with IoT.
- To demarcate what IoT applications need and what distributed ledger technology (DLT) can offer.
- To present the state of the art in the integration of Blockchain with IoT.

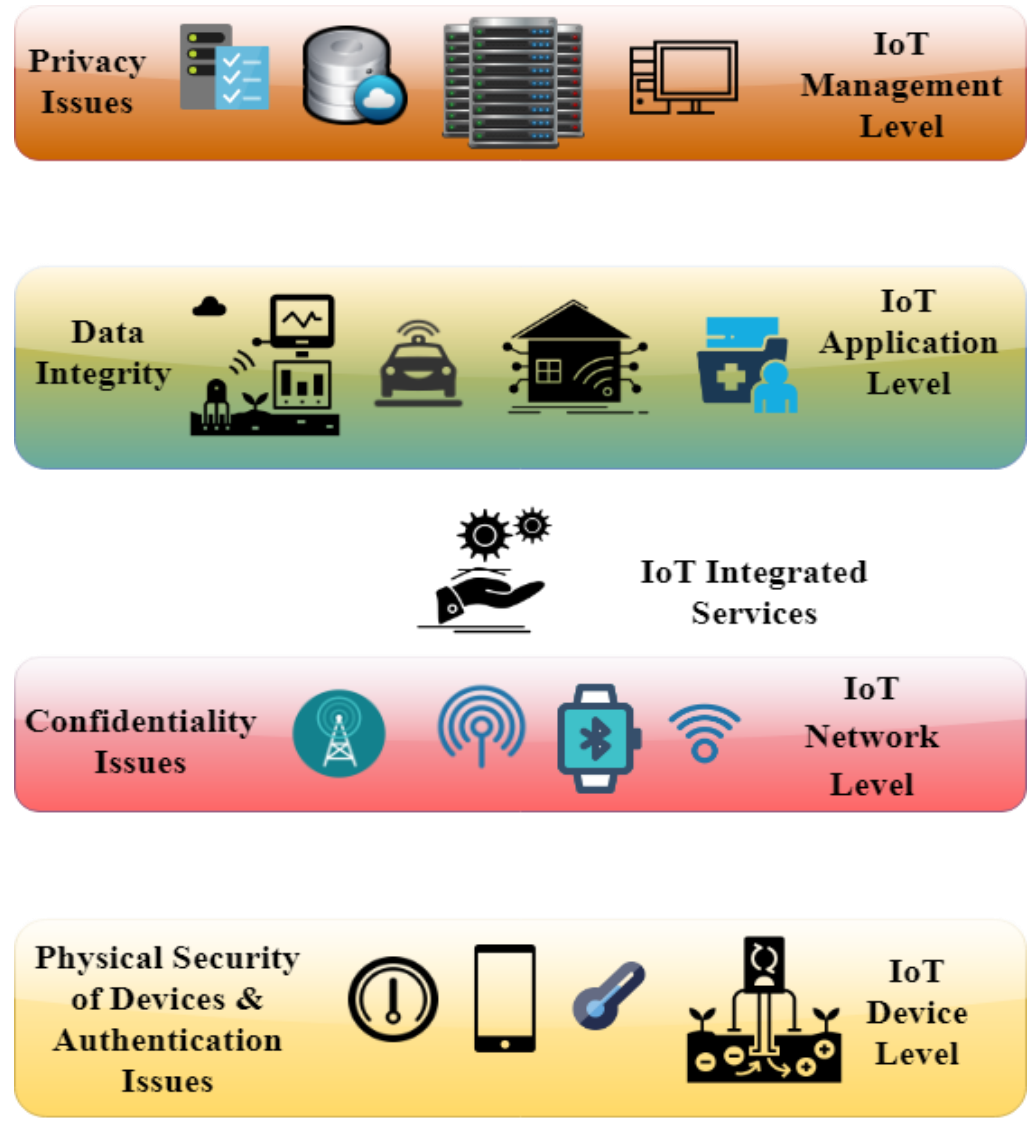

***Figure 1: Security Challenges at different levels of IoT layered architecture***

- To identify and categorize IoT applications based on the issues addressed by Blockchain in the particular application area.

The rest of the paper is organized as follows: Section 2 discusses the security challenges in IoT, Section 3 presents the functioning of Blockchain, and Section 4 describes the need for using Blockchain in IoT. Section 5 discusses state of the art and the existing surveys and research that has gone in integrating Blockchain and IoT Blockchain; it analyses by taking some application domains of IoT where the application of Blockchain has succeeded in bringing trust and security among the different parties involved.

## II. CHALLENGES AND SECURITY THREATS IN THE IoT ECOSYSTEM

Fig. 1 shows the primary security issues at different levels of IoT layered architecture. Safeguarding the device-level security of IoT objects is a cumbersome job considering the whole IoT assets and the possible vulnerabilities they hold. IoT applications should be developed, keeping in mind the design considerations such as storage space, computational power, physical security, communication bandwidth, cost, and latency. The total absence or the lack of regular firmware updates for the installed devices makes it easy for the attackers to exploit the weaknesses in them. The data generated by the devices can be leaked during transmission. The recently instigated Mirai botnet attack, which was one among the biggest distributed denial-of-service (DDoS) attacks in which malicious aggressors intruded into the network by manipulating IoT devices, which were poorly secured, emphasizes the need for strengthening the device-level security in the IoT domain (Fruhlinger, 2018). The low level of security and the high connectivity among the devices made it easy for the attackers to manipulate a few of the devices to obtain the data and make it dysfunctional. A malware named Mirai was installed in the IoT devices using network data from a ”zombie network” through Telnet and elementary dictionary attack procedures. Those devices whose security was compromised became part of the botnet, which was then exploited to continue the DDoS attack.

To provide end-to-end host security at the transport layer, Transport Layer Security (TLS) and Datagram Transport Layer Security (DTLS) can be used, thus ensuring host anonymity. Nevertheless, the problem with solutions like TLS and VPNs is that it proves to be unsuitable for many applications because they are not scalable (Miraz,2019). The existing solutions suggest the usage of the Secure Socket Layer (SSL) protocol encrypts the user’s data while relaying to the cloud storage. This can be used for ensuring secure communication between the devices and the cloud. At the application level, ensuring integrity becomes a significant problem as more and more data gets aggregated from different devices. However, in the IoT management level, as shown in Fig.1, while user-data resides in the cloud servers, there is a chance of it getting shared between or even sold to various companies, violating the user’s rights for privacy and security and further driving public distrust. In most cases, the devices may require the end user’s private data such as name, location, cell number, etc. or social media account information, which can expose a lot about the user to any malicious attacker who gets it. Thus, there is a lack of trust in the level of privacy offered by centralized infrastructures like a cloud. Ferrag et al. (2018) surveys a few privacy-preserving schemes and points out the technical flaws in them. The usage of privacy-aware ids during user-to-device and machine-to-machine communication is suggested in (Elrawy et al. 2018) for avoiding sensitive information tracking. Many researchers have proposed several security risk models and threat models for IoT. In Abdul-Ghani et al. (2018), IoT attacks have been systematically modeled, and the security goals have been analyzed. An attack taxonomy and the corresponding measures to tackle them have been listed. The attacks in the IoT environment have been broadly aligned as Physical-device based attacks, Protocol-based attacks, Data at rest-based attacks, and Application-based attacks. The work by Pongle and Chavan (2015) surveys the attacks on RPL and 6LoWPAN (IPv6 over Low-Power

Wireless Personal Area Networks). In Kouicem et al. (2018), the authors have highlighted the significant challenges of deploying modern cryptographic mechanisms in different IoT applications as mobility, heterogeneity, resource constraints, lack of proper standardization, and scalability. In Elrawy et al. (2018), the authors have surveyed the intrusion detection systems (IDS) for IoT based systems. They claim that IDS can be a potential solution to reducing the network-level attacks on RPL (Routing Protocol for low power and lossy network). It has been shown that IDS can also reduce the chance of a DoS or DDoS attack that aims to disrupt the system by bombarding requests. Bassi et al. (2013) describes the trust factor as a fundamental quality of any IoT system, which depends mainly on the performance in M2M communications and the computational performance and interoperability of the IoT systems. A comprehensive solution to create a secure IoT environment should be an automatic data processing real-time platform that encompasses services like data encryption and authentication, thus promising data integrity, access control, privacy, and it should be scalable and less expensive. The ability to promptly detect and isolate the devices whose security has been compromised is a much-needed requirement to ensure device-level security in any IoT application. The emergence of Blockchain has paved the way for exploring its potential in providing some or all of these security services in the IoT system.

## III. BLOCKCHAIN CONCEPTS

Although numerous papers have elaborated on the techniques and concepts of Blockchain and smart contracts, we must highlight the basics. Fernandez-Caramez and Fraga-Lamas (2018) gives a good description of the blockchain functioning.

A blockchain is a distributed ledger of immutable records that are managed in a peer to peer network. All the nodes in the network ensure the integrity and correctness of the Blockchain through consensus algorithms. This method of placing trust in a network of nodes strengthens the security of the Blockchain. Blockchains can be classified into public, private, or consortium based on whether the membership is permissioned or permissionless. In public blockchains, any user can become a member. There are no membership restrictions, and they are only pseudo-anonymous. Bitcoin and Ethereum are good examples of public blockchains. However, private blockchains are owned and operated by a company or organization, and any user who wants to join the network has to request for membership through the concerned people.

Thus there is only partial immutability offered. Private and consortium blockchains are faster in transaction processing compared to the public counterparts, due to the restricted number of members. Examples of some permissioned blockchains are HyperLedger (Fabric Blummer et al., 2018, Hyperledger, 2017) and R3 Corda (Brown et al., 2016). Some private blockchains impose read restrictions on the data within the blocks. Consortium blockchains are owned and operated by a group of organizations or a private community. Blockchain users use asymmetric key cryptography to sign on transactions. The trust factor maintenance within a distributed ledger technology (DLT), can be attributed to the consensus algorithms and the key desirable properties achieved thenceforth. Wüst and Gervais (2018) gives a good description of these properties. Some of them are Public Verifiability, Transparency, Integrity. The main terminologies in Blockchain are discussed below:

### *A. Terminologies in Blockchain:*

1. ***Blocks:*** The transactions that occur in a peer-to-peer network associated with a blockchain are picked up from a pool of transactions and grouped in a block. Once a transaction has been validated, it cannot be reverted back. Transactions are pseudonymous as they are linked only to the user's public key and not to the real identity of the user. A block may contain several hundreds of transactions. The block-size limits the number of transactions that can be included in a block. Fig.2 shows the general structure of a block in a blockchain. A block consists of the version no., hash of the previous block, the Merkle root tree to trace the transactions in the block, hash of the current block, timestamp and nonce value. A blockchain starts with a genesis block.
2. ***Mining:*** Mining is a process in which the designated nodes in the blockchain network called miners collect transactions from a pool of unprocessed transactions and combine them in a block. In mining, each miner competes to solve an equally difficult computational problem of finding a valid hash value with a particular no. of zeroes that is below a specific target. In Bitcoin mining, the number of zeroes indicates the difficulty of the computation. Many nonce values are tried to arrive at the golden nonce that hashes to a valid hash with the current difficulty level. When a miner arrives at this nonce value, we can say that he has successfully mined a block. This block then gets updated to the chain.
3. ***Consensus:*** The consensus mechanism serves two main purposes, as given in Jesus et al. (2018): block validation and the most extensive chain selection. Proof-of-Work is the consensus algorithm used in Bitcoin Blockchain. The proof-of-stake algorithm is much faster than Proof-of-Work and demands less computational resources. The Ethereum blockchains use a pure proof-of-stake algorithm to ensure consensus. Besides Proof-of-Work, there are other consensus algorithms such as Proof of Byzantine Fault Tolerance (PBFT), proof- of activity, etc. Anwar (2018) presents a consolidated view of the different consensus algorithms. Proof-of-Work is a kind of a signature which indicates that the block has been mined after performing computation with the required difficulty level. This signature can be easily verified by the peers in the network to ensure a block's validity. The longest chain is always selected as the consistent one for appending the new block.
4. ***Smart Contracts:*** They are predefined rules deployed in the Blockchain network that two parties involved in a settlement must agree to priorly. Smart contracts were designed to avoid disagreement, denial, or violation of rules by the parties involved. They are triggered automatically in the Blockchain, on the occurrence of specific events mentioned in the rules.

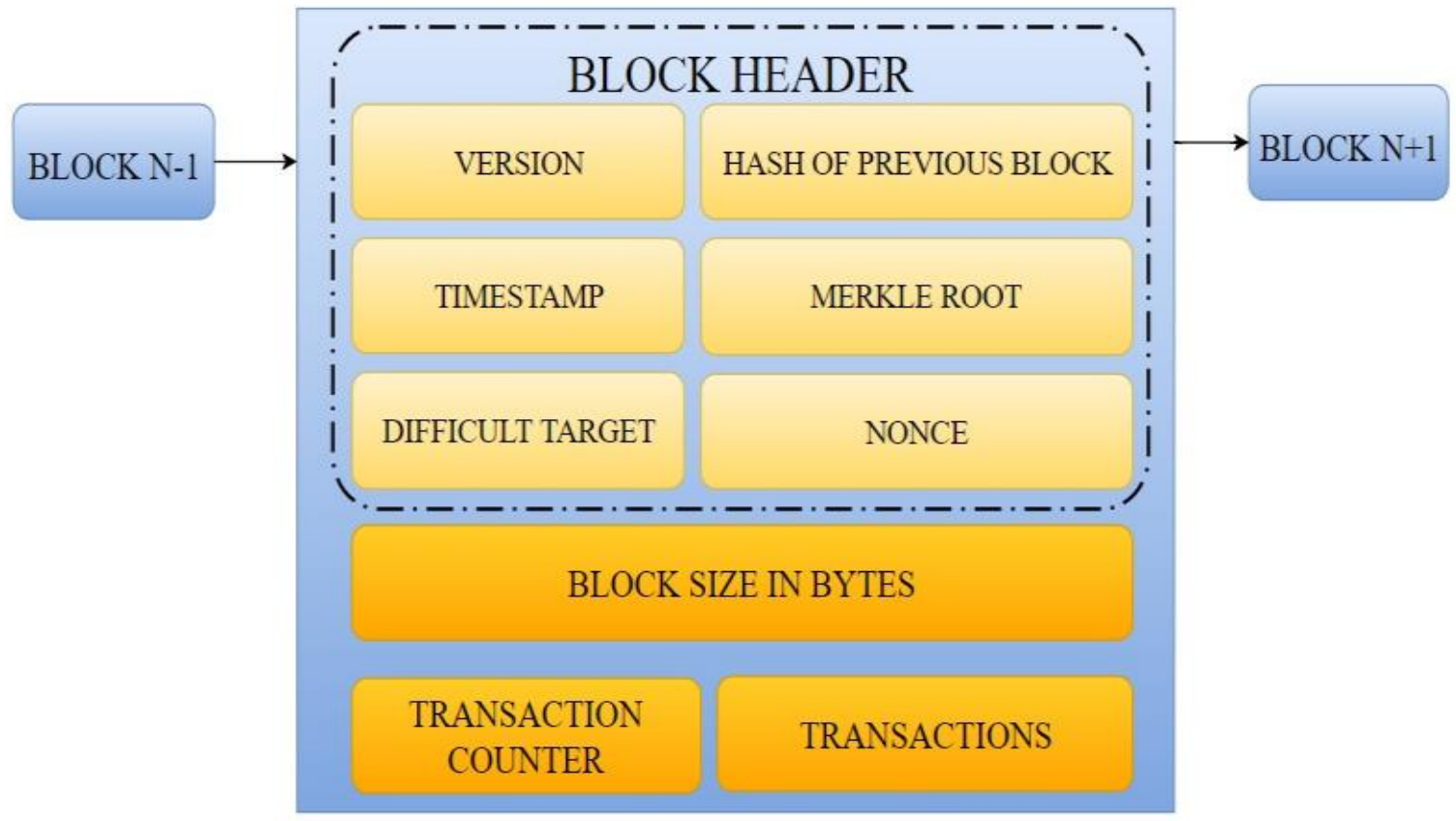

***Figure 2: Structure of a Block in a Blockchain***

*B. Overall functioning:*

Users connect to the Blockchain and initiate a transaction signed with their private key. This transaction is sent to a pool of transactions where it resides until it is fetched into a block by a miner. The miner then generates a new block after gathering transactions from the pool and computing the valid hash of the block. When a miner succeeds in generating a new block, the new block is broadcast to the nodes in the P2P network. All nodes in the network verify the block using a consensus algorithm, and upon successful validation, update it to their copy of the chain, and the transaction attains completion.

## IV. WHY DO WE NEED BLOCKCHAIN IN IOT?

According to Agrawal et al. (2018), the major barricade in the decentralized environment of IoT is the lack of privacy and security of sensitive data that is transferred when connected devices communicate with each other or with the cloud. Data shared cannot be securely encrypted due to the lack of computational power and energy to run encryption algorithms. Thus ensuring data privacy and security is a requirement in any IoT application. The authors in Agrawal et al. (2018) have modeled a security solution using Smart Contracts and Blockchain to ensure continuously secured user-device communications in a smart city and smart home scenario. Using IoT hubs or edge nodes that enable the constrained device to connect to the Blockchain (Agrawal et al., 2018), thus mitigating the necessity to become full nodes, they have tackled the issue of resource and energy constraints. Also, in a distributed environment like blockchain possibility of detecting attacks and restraining further damage is easy since achieving consensus among peers is a necessary condition. Introducing Blockchain in IoT would eliminate the possibility of attacks like DDoS attack, device spoofing, impersonation, injecting malicious code, side-channel attacks, etc. Another reason is that the much-needed security services such as confidentiality, accountability, integrity, and availability come along when we use Blockchain to ensure trust in IoT data. M2M communication between IoT devices can be authenticated. Malicious nodes can be identified and isolated. Recently companies like Chronicle in San Francisco have started using blockchains in their pharmaceutical supply chain for delivering gene therapy drugs. Using a secure IoT platform, they can confirm the quality of drugs and ensure that they don't get damaged or go wrong while in transit. In Miraz (2019) the authors emphasize on the point, "Where IoT is often viewed as a convergence of operational technology (OT) and information technology (IT), Distributed Ledger Technology's (DLT) role as an enabler of the IoT lies in its ability to forge trust, not only at the product level but across an ecosystem of non-trusting constituents."

To determine whether Blockchain is essential in a particular IoT application domain, Ferrag et al. (2018) presents a flow diagram. The consolidated thoughts of it can be briefed as follows: If an IoT application requiring a decentralized security framework demands verification of records between multiple parties who do not trust each other nor a centralized third party and the documents need to be synchronized among these entities, then Blockchain can be applied to arrive at a presumed consensus. However, it should be noted that though the democratic blockchain platform can be used to limit the implications of attack by promoting easier detection, it cannot be taken as an ultimate solution to ensuring IoT security.

Blockchain should not be applied in IoT for merely performing transaction processing or as a substitute for a replicated database requirement. If a traditional database can solve the problem or if all participant trusts a centralized authority, then there's no need for Blockchain in such a scenario. Blockchain should not be used in applications where raw data needs to be gathered in real-time as these incur high latency overheads.

## V. STATE OF THE ART AND EXISTING RESEARCH IN BLOCKCHAIN BASED IOT (BIOT)

Quite a lot of research papers have been written on the convergence of Blockchain with IoT. Blockchain technology, in its current form, cannot be directly absorbed into the IoT domain as most of the IoT applications operate in real-time, and the introduction of Blockchain increases latency, demands high computational power and bandwidth. However, some real implementations though at their nascent level Gundersen (2011) and Hanada et al. (2018) have shown results of a promising future that can be achieved through the convergence of the two technologies. Hence, research continues to delve deep into its use cases, such as the applicability of smart contracts to resolve the majority of the existing business matters like a violation or bypass of policies, forgery, frauds, etc. Many IoT applications may have the requirement of tracking and keeping the participating entities updated on the essential activities that occur in the system. A lot of discrepancies arise in business transactions while procuring products and services, e.g., selling counterfeit products or cheating a manufacturer by demanding payment before the completion of the required service, etc. These discrepancies can be taken care of using sensory data linked with smart contracts. With smart contracts, organizations can safely ascertain funds for current and future operations, the trust factor now being ensured. Bodkhe et al. (2020) survey various consensus algorithms and their applicability in a Cyber-physical system. They also address the challenges and security issues in various CPS application domains.

Dai et al. (2019) discuss the main opportunities that arise in the IoT ecosystem when it is converged with Blockchain, namely: Enhanced interoperability, improved security, traceability, and reliability of IoT data, Autonomic interactions of entities in IoT system. It also describes the architecture of Blockchain of things consisting of data sublayer, network sublayer, consensus sublayer, incentive sublayer, and service sublayer. Makhdoom et al. (2019) discuss methods, like Sharding to reduce transaction processing time, to overcome the challenges associated with the integration of Blockchain and IoT.

Hang and Kim (2019) introduce an IoT platform to assure data integrity of data sensed by IoT devices. It uses HyperLedger Fabric along with a proof of consensus method to achieve consensus. The solution is appropriate to be used in a resource-constrained environment. Rejeb et al. (2019) highlight the emerging research areas that arise with the integration of IoT with Blockchain as Scalability, Security, Immutability, and Auditing, Effectiveness and Efficiency of Information flows, Traceability and Interoperability and Quality

Maroufi et al. (2019) discuss the main challenges in blockchain and IoT integration. Its introduction helps to build a decentralized, scalable, secure, immutable ledger. The heterogeneity of IoT solutions is one of the main challenges involved with the convergence of the two technologies. Panarello et al. (2018) categorize the utility of Blockchain in smart environments into two: Device manipulation and data manipulation. Data manipulation uses Blockchain as a history keeping system, and device manipulation uses smart contracts to make autonomous decisions based on business logic.

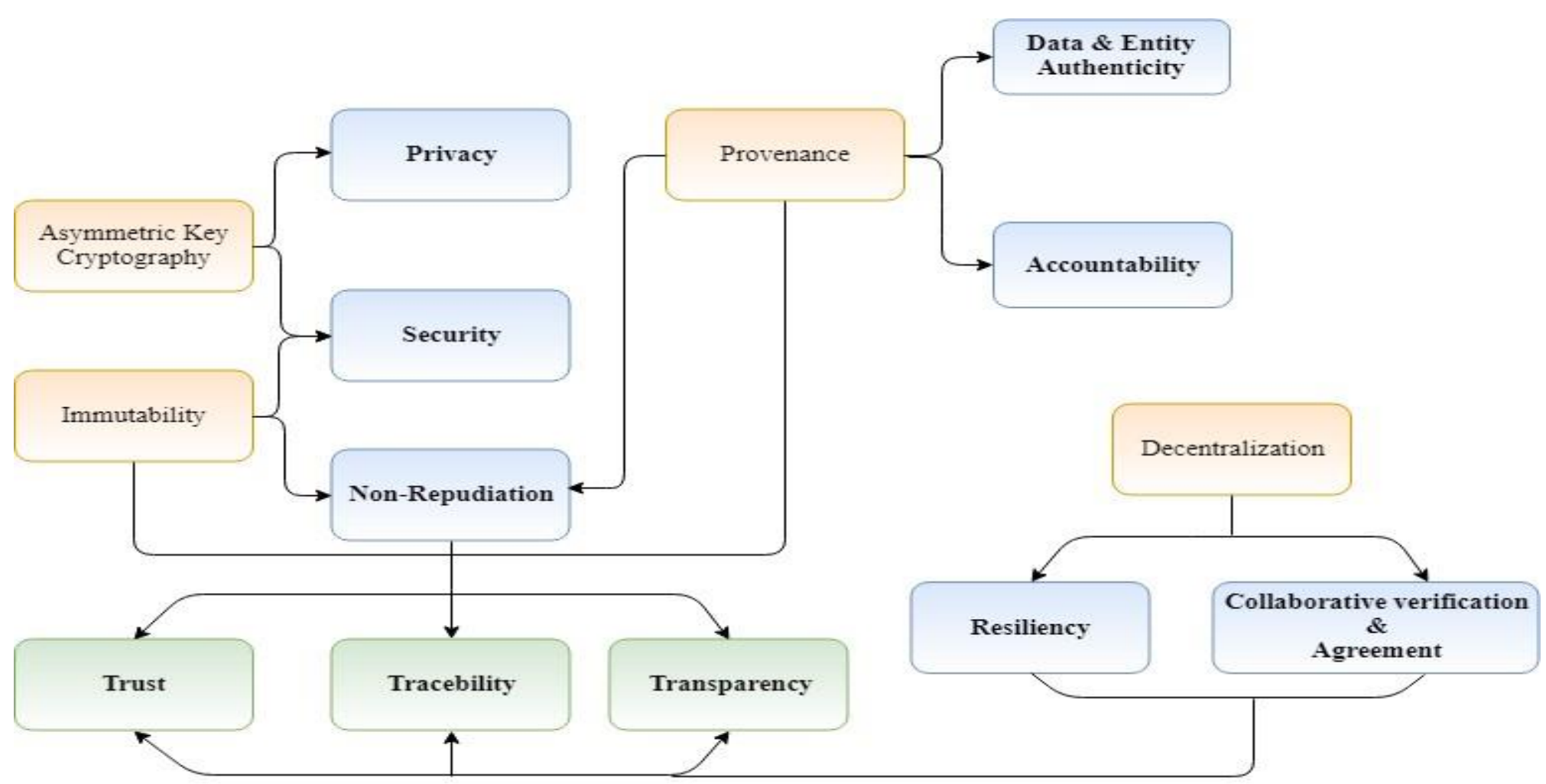

**Figure 3: Relationship between offerings of blockchain and security requirements**

Atlam et al. (2018) describe the comparison between Blockchain and IoT and elaborates on the benefits such as publicity, decentralization, resiliency, security, speed, cost minimization, immutability, and anonymity, of converging the two technologies. Yu et al. (2018) discuss various application domains for the integration of IoT and Blockchain and the main security challenges it solves in an IoT ecosystem. It also gives a view of the Blockchain and IoT integrated framework addressing each layer in the IoT architecture in detail.

Minoli and Occhiogrosso (2018) categorize the different frameworks for IoT blockchain integration into categories such as end-to-end blockchains, Analytics/storage-level, Gateway-level, Site-level, and Device level. A fast payment system for Blockchain backed edge IoT systems is proposed, which ensures security and reliability (Hao et al., 2018). The different phases of the protocol, such as Prepare Phase, Deposit Phase, Fast Payment Phase, Transfer Phase, are described. To prevent double spending problem, a Broker is designated with a pool of deposits. The Broker is responsible for The system ensures security and punishes Brokers. In case they violate the rules of the system. The potential of Blockchain to resolve problems in ensuring trust in the clinical trials in the healthcare industry has been discussed in many papers. Blockchain can not only provide automatized solutions but also establish trust among the different stakeholders such as patients, drug dealers, hospitals, and lab staff and also protect data privacy. According to Benchoufi and Ravaud (2017), Blockchain enables the patients to ensure trust in their private data through access control policies. A proof-of-concept methodology for collecting consent from the concerned people is explained in this context to enable differential privacy. Some of the companies that are currently working on converging the technologies of Blockchain and IoT are:

- Robonomics (Lonshakov et al., 2018): Robonomics platform implements DApp for smart cities and industries
- IoTeX (Dagher and Adhikari, 2018): It is an IoT-oriented blockchain platform that provides application-specific services such as scalability, isolating nodes, deploying applications.
- IoT Chain (Alphand et al.,2018): Uses a lite OS, the Practical Byzantine Fault Tolerance (PBFT), and CPS technology.
- Filament (2018) a startup that's working towards the goal of building a smart economy has their Blockchain solution kit that consists of hardware and software technologies which can be integrated into edge nodes and IoT devices to ensure a secure transaction execution environment (TEE) through Ethereum or HyperLedger blockchains. A hardware plug and play device has been built to enable tracking of events in autonomous vehicles such as vehicle tracking, charging, and making M2M payments. The main functional aspects associated with it are: assigning secret identities to IoT devices, hardware acceleration to reduce computational latency, manage M2M transactions, built-in quality assessment and compliance audit system and key management systems used to connect to Blockchain and obtain Carbon credits, encryption support for ECDSA (Elliptic curve Digital Signature Authentication), ECDH (Elliptic curve Diffie Hellman) and SHA-256 with HMAC option. With this technology, IoT devices can securely communicate valid attested data, which can be verified.
- Xage: A California based cyber-security startup founded in 2016, exploits Blockchain for connecting industrial machines from oil-wells to smart meters. The perk that comes with decentralization is that the security of every device connected will have to be compromised to obtain control of the data in Blockchain. The security of M2M communication at the network and transport layer is secured as they happen over the Blockchain. The Xage Security Fabric provides a comprehensive solution for modern industrial operations. The main features of the fabric include: safeguarding all equipment from the latest IoT devices to vulnerable legacy systems, identity management, single sign-on, and access control. The company provides two new innovative mechanisms to ensure security: hierarchical tree system and super-majority consensus (Xage, 2019).
- Grid+ (Daley, 2019): Using the Ethereum blockchain, it grants consumers permission to access low powered, energy-saving IoT devices.
- Hypr (Daley, 2019): Being a New York-based company, it stores unique biometric login information such as palm, face, eye, a voice in Blockchain, and implemented a DLT digital key to allow homeowners to have single-point access to smart doors and smart entertainment centers.
- ShipChain (Lawrence, 2020): The Company combines a ship and trace platform based on Blockchain with information from near field communication (NFC) tags that observes and shares the temperature of goods throughout the supply chain, thus providing visibility and trust between producers, transporters, and consumers.
- Chronicled (Daley, 2019): The San Francisco based Company focuses on tracing and managing the supply chain for pharmaceuticals and food supply using Blockchain. It uses IoT enabled shipping containers and sensors for this.
- Netobjex (Daley, 2019): The Company created a standard mechanism for M2M communication for IoT devices. It uses IoToken to enable customers in a restaurant to pay their bill, and in drone delivery, it marks the point of delivery and the transaction for payment.
- Some of the companies like BIMCHAIN, Briq, ULedger, Hunter Roberts OEG & Gartner Builders, Probuild, Tata Steel, Lifechain by Costain are working towards the integration of Blockchain & IoT in the construction sector to improve verification of identities and keep track of the progress of their work (Mazhanda, 2019).
- Eight companies namely, Cisco, Asterix, Radiflow, Xage Security, Sumo Logic, BlackRidge Technology, TDi Technologies, Spherical Analytics, have been selected by the National Cybersecurity Center of Excellence (NCCoE) for working towards securing Industrial Internet of Things (IIoT). The main reasons for introducing Blockchain in IIoT is to secure M2M communication, authenticating, and authorizing a transaction.

Ellervee et al. (2017) presents a good comparison of some blockchains and describes the scope of Blockchain, clearly defining what Blockchain can do and what it cannot. Users and block validators or miners represent the actors in the Blockchain. It also describes the processes that occur in the system like network discovery, creating transactions, signing, issuing assets, mining, and assigning permissions, platforms (permissioned or permissionless), and data models. Fig. 3 represents the outline of how we can manipulate Blockchain to ensure security. The figure helps in understanding the relationship between offerings such as immutability, provenance, etc. and which aspect is needed to satisfy the specific security requirements in user applications.

In the following, we present some use-cases, that provide an overview of how Blockchain can be integrated into IoT. We have classified them based on the functional aspects of Blockchain that they utilize and present a sample system model wherever necessary. Some general issues, such as usage of Blockchain for authorization, access control, device identification, data security, secure M2M communication, have been mentioned in various research papers. However, their practical implementation is still in question. Hence these use cases are not discussed in detail.

### *A. Blockchain to ensure Privacy, Security, Non-Repudiation and Identification and Isolation of malicious nodes*

Generally speaking, privacy is a primary concern in any IoT application. Here, we present some major IoT applications where the loss of confidentiality can result in catastrophes and describe how Blockchain can help in avoiding it. We have highlighted the lack of privacy in IoT and described its reasons in detail in Section II.

*1)* ***Intelligent Transport System:*** In an intelligent transport system (ITS), we have vehicles communicating with one another to exchange safety messages and navigation information in real-time such as road conditions, navigation information, traffic jams, accident cases, etc. When vehicles communicate with one another, they share information such as location and user's personal information to identify each other. Hence privacy-preserving schemes are required to protect such sensitive data. Similarly, messages circulated in the network can also be forged. To avoid this and improve trusted communication in VANETs, we look into the applicability of Blockchain in ITS and to what extent it succeeds in solving the problem of Privacy, Security, and Trust.

*a) System Model:* The system model has been explained pertaining to the framework mentioned in Singh and Kim (2017). Here the actors are the vehicles. In this work, Blockchain has been put to use in a vehicular network of autonomous vehicles as a secure data-sharing framework. Here the distrust is between vehicles spreading the information and vehicles receiving this information and using it to perform actions such as taking a lane diversion. Certain vehicles have been designated as Miner nodes based on their willingness to contribute to the network. Consensus achieved in Blockchain helps to ensure trust in the navigation message disseminated and to decide which autonomous vehicle can cross an intersection. The framework implemented has negligible latency associated with it.

Other use-cases of Blockchain in ITS is for vehicle leasing, parts provenance, vehicle tax payment, ride-sharing, performing automatic payments at parking areas and fuel station, or charging station in case of electric vehicles.

The works in Li et al. (2018), Singh and Kim (2017) and Huang et al. (2018) present different methodologies by which Blockchain has been used in VANETS and intelligent transport systems for secure communication and transaction processing. In Singh and Kim (2017), the issues identified in smart vehicle communication are lack of trust, reliability, and data privacy. The authors have introduced a unique identifier called Intelligent Trust Point (IV-TP) similar to a Bitcoin id to ensure privacy in inter-vehicular communication. They have also brought out a proof of driving consensus protocol to maintain consensus. In Yuan and Wang (2016), the authors have elaborated on the design of a blockchain-based framework for ITS. They have presented a seven-layer conceptual model and considered the integration of Blockchain with ITS for creating a parallel transportation management systems (PTMS). A ride-sharing Dapp called La'zooz has also been exemplified. The authors of Lu et al. (2018) investigate the privacy-preserving schemes that can be applied in VANETs to ensure trust and anonymity. They have validated the entity-centric model, which is based on a reputation mechanism to judge the trustworthiness of a vehicle and data-centric model in which trust in the message propagated is computed through various metrics. Finally, they have used a protocol that places faith in a message using a mixture of these two techniques. They have formulated a proof of presence to arrive at a consensus. However, the model proposed by them assumes that the road-side units have adequate computing power. Hanada et al. (2018) studies a decentralized application (DApp) using Blockchain and smart contracts for automatic gasoline purchases in a smart transport system.

They have used smart contracts to counter the problems that affect the aspects of trust and transparency in the services provided by IoT applications.

For making this clearer, let's take the example of a fuel station that delivers its services based on smart contracts and IoT. Consider an interface between the fuel pump and the customer's car. The entities involved in the IoT ecosystem are the sensors that sense the level of fuel in the car and the fuel pump that stops when the fuel is filled. A DApp developed for performing transactions between the IoT nodes can be deployed on the car kit and the fuel pump. The fuel station can then verify payment deposited, and based on this, it can circulate the fuel consumption details to a smart contract. A list of fuel stations based on the user's current location can be checked in the DApp for filling gas.

In Li et al. (2018), CreditCoin, a privacy-preserving incentive announcement network based on Blockchain, has been proposed to report traffic issues and accidents using VANETS. It includes two components: the announcement protocol and incentive mechanism. In this system, the users who complete successful missions are rewarded with credit coins, which actively allows consistent traffic incidents to be reported and verified.

Hîrţan et al. (2020) propose a reputation system that focuses on preserving the privacy of users in an Intelligent Transport System in the cities of San Francisco, Beijing, and Rome. It introduces a consensus mechanism for traffic data sharing, that uses the reputation of the node and the validated answer received from each node in the cluster to achieve consensus in a system of unreliable nodes.

The introduction of Blockchain in ITS helps tackle the privacy concern by introducing anonymity and dynamic heterogeneous key management schemes for ensuring the security of messages in transit.

*2) **Smart Homes and Smart Buildings:***

***System Model:*** A typical home network in smart homes of the future, consists of smart doors and gates, smart lights, smart walls, wearable devices, smart ventilation system, smart home appliances such as refrigerator, AC, Washing machine, etc., smart gardening system, smart surveillance and intrusion detection system (IDS), smart garage, smart washrooms and toilets, smart baby and old people tracker and smart whatnot. These devices communicate with each other and with the user to enhance comforts and living experience. Normally, a smart device in a home network may have shared users, and to personalize the experience, it may be configured with the user profile of its current user, and this data may be shared with other devices for understanding user's choices. However, this collaborative data collection and decision making in the home network comes at the cost of compromising privacy. Hence we use Blockchain to secure the communication between IoT devices. The data in the block consists of the data exchanged between devices, transaction information in case devices perform transactions with the outside world.

The private details of a user can be manipulated by attackers who hack into these devices to perform malicious activities. The stakes are high, including detecting the presence of people at home. Apart from this dangerous compromise in privacy, we also need to take care of the security of devices as they relay information about the location of the smart home for various application purposes. Ensuring device security is another major challenge in a smart home ecosystem. Most vendors do not provide regular security updates for the firmware in smart devices, which leaves vulnerabilities. Hence it is necessary to confirm the security of the connected network of devices installed in a smart home be firmly ascertained. Even more critical is ensuring the security and privacy of devices in smart buildings where numerous employees work.

For this, Dorri et al. (2017) have proposed a scheme using shared keys with a particular lifetime to allow only trusted devices to communicate in a smart home network. Each device in the home network is registered with a home miner. The smart homes are connected to an overlay that enables provide hierarchical and distributed processing. Leiding et al. (2016) presents an access control mechanism that can be applied to Blockchain-based home networks where Security access managers (SAM) are used to identify malicious nodes and prevent them from generating any further transactions in the network. This strategy can be very useful to enable device-level trusted interoperability in smart homes and smart buildings and also in other use-case scenarios such as ITS, in general. Zhou et al. (2018) and Bastos et al. (2018), the authors have used smart contracts and Blockchain to bring about a secure solution for securing communication in a home network.

***3) Smart Healthcare:***

***System Model:*** When we talk about smart healthcare, it is characterized by the vast volume of data from heterogeneous sources varying from wearable devices that track the personal health status of the user to the Doctor's diagnosis and evaluation data and prescriptions. Electronic Health Records are used to integrate and store all this data in digital form. However, since patients may visit different hospitals for different types of diseases and treatments, this data gets fragmented and inconsistent. If we use a central database to store and access this data, it invites security and privacy problems as the data may be accessed or modified by unauthorized parties. Many such cases are stated in papers where the patient data is sold to people in return for money and personal benefits. Hence there's a need for controlling access to such important information to ensure that patients receive proper healthcare services.

Numerous techniques have been proposed on Blockchain to promote a trusted medical data-sharing platform. Most solutions define a publisher-subscriber or patient-requester relationship between the communicating entities involved (Rifi et al., 2018, Shen et al., 2019). In Rifi et al. (2018), the author have proposed and implemented a method using edge servers to overcome the computational power constraint in a smart healthcare data collection network. They have defined the main actors in their network as publishers who are the patients whose health is monitored using sensors and subscribers who require the publisher's data to perform health monitoring and diagnosis. The subscribers who can have access to the health data are added by the publisher to their corresponding address. Thus only subscribers authorized by the publisher gain access to the health data. The patients are monitored within the premises of their smart home. The smart home has a local gateway that connects to an edge server for performing computationally intensive tasks. Using three types of smart contracts, namely: Publisher contract, client contract, and subscriber contract, they have defined the terms and permissions of each actor in the system. For ensuring privacy, the subscriber id and publisher id are verified. In MedChain, a Blockchain-based health data sharing system that uses session keys to safeguard the privacy is introduced (Rifi et al., 2018). In this, a healthcare provider collects data from the sensor devices and performs the mining process. Sessions are used to grant access to the requester of the data.

Ray et al. (2020) discuss Blockchain use cases in healthcare like accessing medical records, EHR claim and bill assessment, Clinical Research, Drug Supply Chain Management, IoBHealth, a dataflow architecture bringing together Blockchain and IoT is proposed, for accessing, storing and managing e-healthcare (EHR) data. Uddin et al. (2020) propose a three-layer architecture for leveraging Blockchain in eHealth services consisting of Sensing Layer, Edge processing layer, Cloud layer. A patient agent (PA) software implements a lightweight blockchain consensus mechanism and uses a task offloading algorithm for preserving privacy.

***4) Smart Manufacturing and Industrial IoT:***

***System Model:*** The evolution of Industrial IoT (IIoT) and smart cognitive and autonomous robots, augmented reality technologies in the manufacturing sector has led to the emergence of the 4th Industrial revolution. This has created a huge impact on the manufacturing sector. Smart factories are becoming a reality with companies like Ubirch bringing automatization through smart strategies like a machine as a service, predictive maintenance, and verification of the condition of manufactured products. Smart devices aid manufacturing in design, manufacture maintenance, and supply of products. With IoT, the factory can be transformed into an ecosystem where everything is connected, well-monitored, secured and monetized. The advantages of this connected environment are: the quality of products can be ensured, and machines and services can be shared. In this shared, connected ecosystem, when machines engage in M2M communication, trust, security, and privacy become at stake. Hackers can target the surveillance or the machines for disrupting the factory operations.

Blockchain can be used to perform trusted communication and ensure privacy and security in the transactions performed in distributed manufacturing. Xu et al. (2019) and Liu et al. (2019) describe the security frameworks developed using Blockchain for IIoT. In Xu et al. (2019), a Blockchain-based Service Scheme is presented. A non-repudiation scheme is proposed, which works based on a homomorphic hash function to ensure whether a particular service promised is delivered or not. In Wan et al. (2019), presented a Blockchain-based security framework for ensuring privacy and security in a smart factory. They have given a layered division to the framework as the Sensing layer, Firmware layer, Management hub layer, Storage layer, and application layer. They have used a blacklist- whitelist mechanism to track and isolate malicious nodes and ensure privacy. They have shown how a lightweight private blockchain can be used to trace the operations of machines and M2M communications. They have combined two models: Bell-La Padula (BLP) model and Biba model to build the data security model using a dynamic identity verification mechanism.

*B. Blockchain for Provenance, Accountability, Traceability, and Transparency*

*1)* ***Blockchain for Smart Grid:*** Blockchains are used in Smart-Grid networks for secure trading of sustainable energy. The evolution of micro-grids and smart-grids have enabled small-scale production and selling of power. Provenance and accountability are the guiding factors that need to be brought into the transactions in the energy-trade microgrid network for efficient interoperability.

***System Model:*** The working of a smart-grid network in a city can be explained like this. Excess renewable energy such as solar energy or wind energy produced at homes can be stored in batteries, which can later be sold to other homes in the neighborhood or companies. A smart-meter installed keeps track of the energy units produced or consumed in real-time and relays this information to the energy distributor for monitoring and billing. With smart-grids becoming a reality, there is a need for a secure energy-trading platform to establish trust and track transactions that occur in the smart grid network.

The state of the art of applying Blockchain to micro-grids and smart-grids brings us to the Brooklyn Micro-grid solutions offered by LO3 Energy (Gundersen, 2011). The smart-contracts linked to the local grid are used to tokenize the extra energy and decide where to buy the electric power from. The energy sharing market consists of four main concepts: Tokenization, P2P Markets, Prosumers, and Community Micro-grids, and Energy-service Companies (CMESC). Tokenization involves converting surplus energy to energy credits. The P2P markets are a network of nodes established in the smart-grid between neighbors and prosumers to energy-producing companies. Prosumers are called so because of the dual role they play as producers and consumers in the energy trading system. In Pieroni et al. (2018), Blockchain is applied for securing communication, transactions, and tokenization of energy. A Smart energy blockchain application prototype has been implemented in the work that connects to the Blockchain in the P2P network. Li et al. (2017) has given a good demonstration of the smart-grid detailing the entities involved as Smart meters, energy nodes, and energy aggregators. Using a consortium blockchain, a unified blockchain for the energy credits transferred has been presented, which can be deployed in the Industrial IoT ecosystem. This enables the user to trace their transaction history and transact with trusted prosumers or companies efficiently. Daghmehchi Firoozjaei et al.(2020) propose a privacy-preserving hybrid blockchain framework called Hy-Bridge, which uses subnetworks that preserve the privacy of IoT users in a microgrid. It separates transactions of the power grid from those of the micro grid. It has three layers IoT user layer, platform layer, and enterprise layer.

*2)* ***Smart Cities:*** A waste management system for a future smart city is presented in Lamichhane et al. (2017), which is characterized by smart garbage bins(SGB) involving the sensors that detect and alerts the garbage collection service when the bin is full, actuators and smart locks to lock the SGBs. The different entities involved in the system are citizens, Municipality, Waste management operators who dispatch driverless vehicles that perform garbage collection, recyclers who recycle the waste to obtain rewards. Blockchain can be used to record operations and trace the transactions associated with garbage collection and recycling. In Nagothu et al. (2018), collaborative smart surveillance has been presented which uses access control rules specified in the smart contracts to restrict access of unauthorized nodes at fog level to connect to a cloud node containing Blockchain that stores surveillance data in various places in a smart city. In Nikouei et al. (2018), the authors have proposed an authentication scheme integrated into the Blockchain is used to share and aggregate the surveillance data in two trustless domains at the edge and fog level.

| Sl No. | IoT Application Domains | What Blockchain Offers | | | | | | | | |
|---|---|---|---|---|---|---|---|---|---|---|
| | | Identification & Isolation Of Malicious Nodes | Provenance | Security | Privacy | Traceability | Trust | Transparency | Non-Repudiation | Accountability |
| 1 | Smart city | ✓ | ✓ | ✓ | | | ✓ | ✓ | | |
| 2 | Smart Grid | | ✓ | | | ✓ | ✓ | ✓ | ✓ | ✓ |
| 3 | Smart Healthcare | | ✓ | ✓ | ✓ | ✓ | ✓ | ✓ | ✓ | ✓ |
| 4 | Supply chain management | | ✓ | ✓ | | ✓ | ✓ | ✓ | ✓ | ✓ |
| 5 | Intelligent Transport Systems | ✓ | ✓ | ✓ | ✓ | | ✓ | ✓ | ✓ | ✓ |
| 6 | Smart homes & Buildings | ✓ | ✓ | | ✓ | | ✓ | | ✓ | |
| 7 | Industrial IoT & Smart Manufacturing | | ✓ | | | | ✓ | | ✓ | |

***Table 1: Solutions Offered by Blockchain in Various Applications in IoT***

### *3) Smart Supply Chain Management:*

***System Model:*** A smart supply chain consists of smart objects which have a unique identity and are connected through wireless technologies. They can store information about their conditions like temperature, humidity, etc. and communicate to the Internet through wireless technologies. Then there are suppliers, distributors, transportation services, and consumers. In a conventional supply chain environment, the condition of the products is not traced at each checkpoint. Hence huge loss is encountered by the buyers. Sometimes products transported get tampered or replaced. Thus to ensure the security of manufactured products throughout the supply chain, we need an automated decentralized system that helps to track and assess the quality of product real-time.

Blockchain has been employed in a food supply chain system to overcome this problem. In Lin et al. (2018), Blockchain-based food traceability is proposed. This system enables to ensure food safety by tracking the food product and storing its condition at each stage. Through smart contracts which it establishes between buyer and seller, it provides a transparent system to the buyer, which allows them to verify and ensure the quality of the food they buy. Blockchain and IoT sensors together provide both parties provenance, traceability, trust, and transparency. Through this system, the seller can also identify at which stage in the supply chain the product suffered damage so that they can take remedial methods to avoid the same in the future. They can also determine whether the product has been tampered by scanning the RFID and tracing the sensor values collected, which are stored as metadata in the blocks. Table I summarizes the most relevant security services offered by Blockchain in each IoT application discussed so far.

## VI. FUTURE WORK

The applications discussed in the previous section, depict a bright future for the convergence of Blockchain and IoT. However, the inclusion of Blockchain always introduces the probability of a 51% vulnerability. The papers surveyed predict this as far from the possible case. Lightweight algorithms being researched extensively might help in tackling energy and power constraints in IoT environments. However, there’s a need for developing a unified blockchain framework for IoT applications. Also, though many initial research works claim that Blockchain is impregnable, there have been many incidents of breach of security of Blockchain like key tampering attacks, Sybil attack, etc. which have been elaborated in (Ferrag et al. 2018). Since the Blockchain in its current form cannot guarantee ultimate security, research needs to focus on identification and resolution of flaws so that developers can build more successful secure versions of Blockchain, which, when integrated with IoT, can provide a secure and safe environment for the application users.

## VI. CONCLUSIONS

The IoT applications face a lot of challenges in terms of security, which can be solved using Blockchain. We have iscussed the major security issues in IoT and exemplified how the application of Blockchain can help in addressing these issues. Many research texts that were surveyed have applied Blockchain to IoT applications without actually ensuring whether it is necessary. Careful assessment needs to be done before applying Blockchain for a particular application. We have analyzed the need for using Blockchain in a particular application area. We have presented the decision criteria and mentioned where it can be used and what offerings it can provide and what it cannot. Also, we have explained a general system model and supportive use cases that can help in the integration and development of Blockchain-based secure IoT solutions.